\documentclass[prb]{revtex4}
\usepackage{graphicx}

\begin{document}
\title{Energetics and Stability of Nanostructured Amorphous Carbon}
\author{M.G. Fyta, I.N. Remediakis, and P.C. Kelires}
\affiliation{Physics Department University of Crete, P.O. Box 2208, 
710 03, Heraclion, Crete, Greece}

\begin{abstract}

Monte Carlo simulations, supplemented by {\it ab initio} calculations, 
shed light into the energetics and thermodynamic stability of 
nanostructured amorphous carbon. The interaction of the embedded 
nanocrystals with the host amorphous matrix is shown to determine in a 
large degree the stability and the relative energy differences among 
carbon phases. Diamonds are stable structures in matrices with sp$^3$ 
fraction over 60\%. Schwarzites are stable in low-coordinated networks. 
Other sp$^2$-bonded structures are metastable. 

\end{abstract}

\maketitle

\section{INTRODUCTION}
\label{sec:intro}

Interest on nanostructured amorphous carbon (na-C) is steadily
growing for its properties, both mechanical and electronic,
that will supplement those of the traditional, single-phase a-C.
Na-C can be described as a composite material in which 
carbon nanocrystallites, of various sizes and phases, are embedded
in the a-C matrix. This novel hybrid form of carbon
offers the unique possibility to intermingle the properties of 
carbon nanostructures\cite{reviews} with those of pure a-C.\cite{Rober,Silva}
For example, since some of these nanostructures are proposed to be
insulating, while others to be metallic, the possibility is opened
for tailoring the electronic properties of a-C by
controlling the type and size of the embedded structures.

Na-C films can be composed by a variety of
carbon nanostructures, ranging from diamond crystallites, to 
concentric-shell graphitic onions\cite{reviews} bound by van der Waals 
(VDW) forces, to entirely three-dimensional sp$^2$ covalent conformations 
with no VDW bonding. The latter include porous, open graphene 
structures with negative curvature (schwarzites). 
\cite{Vanderbilt,Lenosky92,Keeffe,Winkler} 
Synthesis of na-C films can be achieved by different methods such as
arc plasma methods,\cite{Amaratunga} leading to networks with 
bucky onions, or by supersonic cluster beam deposition (SCBD) 
techniques,\cite{Milani97,Donadio} producing schwarzites, or by electron 
irradiation of a-C, producing onions and eventually nucleation of diamond 
cores.\cite{Banhart}

Significant insight was gained through these experiments but several 
issues about the structure, energetics and stability of na-C are 
not yet clear. Theoretical work up to now has been mainly 
focused on the free-standing periodic nanostructures 
themselves.\cite{Vanderbilt,Lenosky92,Keeffe,Winkler} The transformation 
from graphitic to diamond nanostructures under irradiation conditions 
is also well studied.\cite{Banhart} However, knowledge about 
{\it the interaction of the embedded nanostructures with the host 
amorphous matrix} is very limited. This interaction is a crucial factor 
since it determines the stability of a given nanostructure. Variations 
of the density and coordination of the amorphous matrix are expected 
to control the stability and the relative energy differences among the 
carbon phases.

Here, we propose a novel theoretical approach to the problem
of energetics and stability of na-C, which aims at unraveling the 
fundamental principles governing the interaction of nanostructures with 
the a-C matrix. It is based on Monte Carlo (MC) simulations within 
the empirical potential method. This approach is not limited to a-C, but 
can be applied to any nanostructured material. It is a general method, no
matter which energy functional is used. For each computer-generated 
nanostructure embedded in the amorphous matrix, with varying coordination 
$z_{am}$, its formation energy $E_{form}$, interfacial geometry, density, 
and stress state were calculated. Several interesting conclusions 
follow this analysis. For example, the question whether diamonds and 
schwarzites are stable in the amorphous matrix 
is a crucial one and of wide interest, and we give answer to that for the 
first time.

In the following, we first describe our theoretical 
methods. We then explain how we generate periodic cells with various
nanostructures embedded in a-C matrices, we analyze the energetics 
by means of their formation energy, we study their stability as a function
of size, temperature and coordination of the matrix, and we consider the 
possibility of transformations from one nanophase to another. 
Finally, we give our conclusions and prospects for future work.

\section{METHODOLOGY}
\label{sec:method}

The simulations have been carried out using the empirical potential of 
Tersoff,\cite{Tersoff1} which provides a fairly good description of the
structure and energetics of a wide range of a-C phases.
\cite{Kelires94,Kelires2000} 
It was recently used in a similar context by Donadio {\it et al.}
\cite{Donadio} in molecular dynamics simulations of the growth of SCBD 
films. It was nicely shown that pore structures resembling schwarzites 
can be formed. The important question, to which we answer below, is whether 
these are stable or metastable, and at what conditions. 

A limitation in the model is the absence of dispersion (VDW) forces, 
similar to more refined tight-binding or first-principles methods, which 
makes the study of structures involving VDW bonding not possible.
There also is an absence of repulsion between non-bonded $\pi$ 
orbitals\cite{Stephan,Kaukonen,Jager} leading to the overestimation of 
density for a given sp$^3$ content.\cite{Kelires2000,Mathiou} This mostly 
occurs at intermediate sp$^3$ contents, while low-sp$^3$ (evaporated a-C) and 
high-sp$^3$ (ta-C) networks, are properly described in this respect. 
For ta-C (sp$^3$ fraction about 80\%) the overestimation
is $\sim$ 3-4\%. We do not expect this shortcoming to seriously affect
the energetics of the amorphous matrix, because this is determined 
primarily by the $\sigma$ bond character of the network, which is
very well treated by the potential. The absence of $\pi$ repulsion 
has only a secondary effect on the energetics, through the generation of 
denser networks at intermediate contents, not serious enough to change
the trends reported here. 

Since the system under study is a composite material, and
crystalline nanograins are part of it, it is necessary to show
that the empirical potential is adequate in describing these phases.
So, in order to establish the reliability of the potential for the present
problem, we calibrated and checked it with respect to the energetics of 
various crystalline carbon structures by comparing to {\it ab initio} 
calculations. For the latter, we used the Vienna Ab-initio
Simulation Package (VASP) code.\cite{Kresse} We selected a number of 
structures that are each one of them {\it representative 
of a specific group of conformations}, and which are subsequently
embbeded in amorphous matrices: (a) Diamond (D); (b) the BC8 structure for 
carbon (internal parameter $x$=0.0937), fourfold-coordinated with high 
density, a high-pressure phase; (c) polybenzene (6.8$^2D$), a 
threefold-coordinated structure with negative curvature, and with a $D$ 
periodic minimal surface, that can be described as a three-dimensional 
linkage of benzenelike rings\cite{Keeffe}; (d) another sp$^2$-bonded, 
negative-curvature structure, 6.8$^2P$, with a $P$ minimal surface, 
described as a condensation of truncated-icosahedral C$_{60}$ 
molecules\cite{Keeffe}; (e) PCCM (porous conducting carbon 
modification), an sp$^2$-bonded structure with eightfold rings, described 
as a dense packing of compressed nanotubes with shared 
walls\cite{Winkler}; and (f) C$_{168}$, the low-density, 
negative-curvature schwarzite structure (``buckygym'') proposed by 
Vanderbilt and Tersoff.\cite{Vanderbilt} Structures (c)-(f) do not 
involve VDW bonding. Onionlike shell structures bound by VDW forces can 
not be treated with the present potential. However, Polybenzene, 6.8$^2P$, 
and PCCM can be considered as candidate structures resulting from
fragmentation and distortion of embedded fullerenes and
nanotubes in amorphous matrices. There is extensive experimental
research on this method of producing composite carbon films.
\cite{Alexandrou,Hirono,Kozlov,Brazhkin}

The energies of these structures relative to diamond as a function of 
the density are given in Fig. 1. Our {\it ab initio} energies for 
6.8$^2D$, 6.8$^2P$, BC8, and PCCM are consistent with previous 
results.\cite{Keeffe,Winkler} Overall, there is very good agreement
between the empirical and the {\it ab initio} calculations both with
respect to the energies and the density of the structures, providing
confidence in using this potential for the investigation of na-C. The
agreement is particularly good for the C$_{168}$ schwarzite. Only in
the case of PCCM a noticeable relative energy difference exists, 
but it does not affect the trends in our investigations. The lowest 
energy structure is polybenzene, followed closely by PCCM and C$_{168}$.
The highest energy structure is BC8, as expected, since it is a
high-pressure phase.

\section{RESULTS AND DISCUSSION}

\subsection{Generation of structures}

The embedded nanostructures are formed by MC melting and subsequent 
quenching at constant volume the corresponding crystal structures,
while keeping a certain number of atoms in the central portion 
of the cells frozen in their ideal crystal positions. 
Periodic boundary conditions are applied to the cells. The total
number of atoms ranges from $\sim$ 2500 to 4096, depending on the
structure, while the number of atoms in the nanocrystals ranges from
$\sim$ 300 to 500.

After quenching, which produces amorphization of the 
surrounding matrix, the cells are thoroughly relaxed with respect to 
atom positions and density. Relaxations are particularly important at the 
interface region, where the crystallites mainly adjust to the host 
environment. Cells with varying coordination (density) of the 
amorphous matrix can be formed by changing the initial starting 
density (volume) of the crystal structures. The size (radius ) of the 
nanocrystals is controlled by the choice of the number of the shells kept 
frozen during quenching. Their shape depends on the particular crystal 
symmetry.

Four of the generated composite cells with embedded nanostructures are 
illustrated in Fig. 2. In these snapshots, 
diamond is embedded in a highly tetrahedral matrix ($z_{am} \simeq$ 3.8),
polybenzene and PCCM are embedded in a typical a-C matrix 
($z_{am} \simeq$ 3.2), and C$_{168}$ schwarzite is embedded in a 
low-density matrix ($z_{am}\simeq$ 2.8). In the latter cell, the 
amorphous matrix exhibits a characteristic open nanoporous structure 
resembling random pore schwarzites.\cite{Donadio,Townsend} In the 
polybenzene and PCCM cells, the matrix is more or less homogeneous, in
this sense, but characteristic benzenelike and eightfold rings, 
respectively, can be easily identified. In the diamond cell, it is
interesting to note that there is a remarkable tendency of sp$^3$ atoms in
the matrix to gather and enrich the interface region around the
nanodiamond. A careful inspection of the local structure reveals an
increasing degree of crystallinity, indicating that such tetrahedral
atoms may act as possible nucleation centers for expanding the
size of the diamond nanocrystal under the appropriate conditions.
We discuss this further below.

\subsection{Formation energies}

The central issue of interest in this paper concerns the energetics and
stability of the nanostructures. Whether a certain nanocrystal is stable
or not can not be told by just comparing the total energy of the
composite structure to a reference energy, e.g. diamond. One needs to
consider the interaction of the embedded configuration with the host.
Here, this is taken care of properly by defining the {\it formation energy} 
of a nanocrystal, given by
\begin{equation}
E_{form} = E_{total} - N_{a}E_{a} - N_{c}E_{c},
\label{form}
\end{equation}
where $E_{total}$ is the total cohesive energy of the composite system
(amorphous matrix plus nanocrystal), calculated directly from the
simulation, $E_{c}$ is the cohesive energy per atom of the respective 
crystalline phase, $N_{c}$ is the number
of atoms in the nanocrystal, $N_{a}$ is the number of atoms in the
amorphous matrix, and $E_{a}$ is the cohesive energy per atom of the
pure, undistorted amorphous phase (without the nanocrystal) with
coordination $z_{am}$. A negative value of $E_{form}$ denotes stability
of the nanostructure, a positive value indicates metastability.

To compute $E_{form}$, we need $E_{a}$ for an arbitrary $z_{am}$.
This is obtained in the following way. We carried out an extensive
energetics study of a-C as a function of coordination, or sp$^3$ fraction, 
over the whole region where a-C is a rigid material. Such a study has not 
been done before. The results of calculations of the cohesive energy of a 
number of a-C configurations are plotted in Fig. 3. The minimum of energy 
is, not surprisingly, at $z_{am}$ = 3, where predominance of sp$^2$ bonding
is taking place. With increasing coordination, 
$E_{a}$ becomes higher because the mixed (sp$^3$/sp$^2$) phase is 
strained.\cite{Kelires94,Kelires2000} This is maximized in the tetrahedral
region. On the other side, characterized by configurations with many
twofold atoms, the variation is more steep with the maximum energy
occuring when the networks become floppy ($z_{am} \simeq$ 2.4).
\cite{Thorpe} A cubic polynomial best fits the $z_{am} \ge 3$ region, 
while a linear fit applies to the $z_{am} < 3$ region. From these fits 
one can extract $E_{a}$ for any arbitrary value of $z_{am}$. 

We remind the reader that the absence of $\pi$ repulsion in the potential 
generates somewhat denser networks in the intermediate region 
(3.3 $< z_{am} <$ 3.6). This, in turn, overestimates the cohesive 
energies. Correction of this would slightly shift the energies in 
this region to higher values, producing a straight line in Fig. 3, but 
the trends reported here, based on relative energy differences,
remain unaltered.

For each one of the six nanostructures considered in this work,
we have generated a series of cells with different $z_{am}$ of the
amorphous matrix, but keeping the size of the nanocrystal constant. 
Then, application of the above methodology yielded
$E_{form}$ as a function of $z_{am}$. The energy curves 
resulting from this analysis are shown in Fig. 4. There are a number 
of important aspects of these results. The first concerns the question 
of thermodynamic stability. A stable structure has, in principle, the
potential to expand against the surrounding matrix, provided that
the barriers for this transformation can be overcome.
Of the six nanostructures considered, 
only diamond and the C$_{168}$ schwarzite are actually stable 
structures ($E_{form} < 0$), the former at high $z_{am}$ 
($\rho_{am} \geq 3\ gcm^{-3}$), the latter at low $z_{am}$ 
($\rho_{am} \leq 1.2\ gcm^{-3}$). With increasing (decreasing) $z_{am}$ 
diamonds (C$_{168}$) become more stable, respectively. Yet, 
diamonds are overall relatively more stable than schwarzites. The enhanced
stability of diamonds in highly tetrahedral networks indicates that once 
nucleation centers are formed, possibly by intense ion irradiation, 
large diamond regions can be further grown.

The other nanostructures studied in this work are metastable 
($E_{form} > 0$) through the whole coordination range. However, their curves
exhibit a well defined local minimum at which $E_{form}$ is quite small, 
suggesting that their synthesis is possible, and that they can be
maintained in the amorphous matrix under moderate conditions of
temperature and pressure.

Remarkably, there is a reversal of relative stability between these
embedded metastable phases, contrasted to their free-standing
energetics. Take for example the case of the BC8
structure, which initially is the highest in energy (see Fig. 1) but 
when embedded in a matrix with $z_{am} \simeq$ 3.4 becomes only slightly 
metastable. This demonstrates the significance of the interaction between 
the crystallite and the host. 
The interaction becomes favorable when the amorphous environment enforces
relaxations in the BC8 nanocrystal that shift the bond lengths and angles 
to more tetrahedral values. Specifically, there are two sets of bond 
lengths and angles in the BC8 structure: three bonds are 1.62 \AA\, long 
and one is 1.44 \AA\, long (relaxed with the present potential). 
The angle subtended
by a long and a short bond is 101.5$^{\circ}$, and by two long bonds
116.1$^{\circ}$. The embedding induces a shift of the short bond to
1.52 \AA, while the longer bond remains unaltered, and of the small
angle to 109$^{\circ}$ (the large angle remains the same).
A similar reversal in relative stability occurs between polybenzene 
and PCCM. The latter becomes more stable in the amorphous matrix.

The transition of diamonds from metastability to stability 
at $z_{am} \simeq 3.6$ gives us the opportunity to furnish a 
quantitative definition of tetrahedral a-C (ta-C), vaguely referred to 
as the form of a-C with a high fraction of sp$^3$ bonding. 
We define ta-C as the form of a-C with a fraction of sp$^3$ sites 
above 60\%, {\it in which diamond crystallites are stable}. In other
words, the predominantly tetrahedral amorphous network of ta-C is able to
sustain its crystalline counterpart. Networks with sp$^3$ fractions
below 60\% do not belong to the class of ta-C materials, because they
can not be transformed into a stable nanocrystalline state.

One way of checking the stability of nanocrystals, besides referring to 
$E_{form}$, is to subject them to thermal annealing. A stable structure 
should be sustained in the amorphous matrix, while a metastable structure 
should shrink in favor of the host. We quantitatively demonstrate this 
by means of the tetrahedral vector $\vec{v}_{t}$, which is the sum of 
the vectors pointing from an atom to its nearest neighbors, 
and shows the deviation from ideal tetrahedral geometry. 
The magnitude of this quantity is plotted in Fig. 5 for diamond 
crystallites, in both the stable and metastable regions, ``as-grown'' 
and after annealing. The nanocrystals have a radius 
of $\sim$ 8.5 \AA. Already before annealing, the metastable nanocrystal
is heavily deformed in the outer regions near the interface 
with the amorphous matrix. The stable one is only slightly 
deformed. Upon annealing, it is evident that the former structure 
extensively shrinks, and only a small core remains intact, while the
stable structure retains its tetrahedral geometry. 

In principle, a stable nanocrystal not only retains its shape and 
structure upon annealing but, as said earlier, should expand against
the amorphous matrix. We have not seen this to happen dynamically in the 
simulations, which were carried out at relatively high temperatures
(1500-2000 K) to accelerate the deterioration of the unstable
nanocrystals. Actually, such a thermal annealing experiment, either in
the laboratory or on the computer, should take place at temperatures below 
1200 K, i.e., in the region where the tetrahedral amorphous network is 
stable against the transformation of sp$^3$ to sp$^2$ 
bonding.\cite{Kelires94} It is doubtful that growth of diamonds will be 
observed dynamically in simulations at these temperatures because of the 
enormous computational time needed, but we expect this to happen 
experimentally.

\subsection{Phase transformations}

Another interesting aspect of our results is that the formation
energies of 6.8$^2D$ and PCCM at their minimum (see Fig. 4) are only
slightly lower than diamond's at the same $z_{am}$. This suggests the 
possibility of ready transformation of sp$^2$-bonded structures to diamond 
under the appropriate conditions of pressure. To check whether such 
conditions exist, we calculated the stress fields within the embedded 
nanocrystals. We found high compressive atomic stresses\cite{Kelires94} 
in both 6.8$^2D$ and PCCM that approach 50 GPa in the center, even
at their minimum energy configurations. This is illustrated in Fig. 6
for both cases. The compressive stress reduces as we move radially 
outwards to the interface. (For 6.8$^2D$, it remains high up to two thirds
of the nanocrystal radius). The origin of the compression is attributed to 
the lower density and the more open configurations of these nanocrystals
compared to the amorphous background. On the other hand, diamond 
nanocrystals in a ta-C matrix are somewhat denser than the tetrahedral 
amorphous network around them, and they are found to be under 
slight tensile stress.

Such huge compressive stresses in the center of the sp$^2$-bonded 
structures are proper conditions to locally instigate the 
sp$^2 \rightarrow$ sp$^3$ transition and the nucleation of diamond cores, 
provided that the barriers for the transformation are overcome by means of 
external pressure. (Compression is the favorable stress state for the 
formation of sp$^3$ local geometries.)\cite{Kelires94} Let us emphasize 
that the so-produced diamonds would be 
metastable, because their formation energies in this region are positive.
Similar compressive conditions are thought to be
responsible for the nucleation of diamond cores in the center of
spherical onions under electron irradiation.\cite{Banhart} 

\subsection{Size dependence}

Finally, we checked the stability of the embedded nanocrystals with
respect to their size, while keeping the same $z_{am}$ for the 
amorphous matrix. An example of this analysis is shown in Fig. 7
for diamonds. The difference between a stable and a metastable
nanocrystal is striking. The former has a deep minimum bounded
by a steep branch at small radii, indicating a {\it critical size} of
$\sim$ 6.5 \AA, and by a barrier on the right that separates it from other 
less stable states. Note that, due to the periodic boundary conditions
applied to the cells, each crystallite interacts with its neighboring
images. This interaction is attractive for certain sizes (distances),
e.g. for the minimum and the other low energy states, and repulsive 
for other (barrier). On the other hand, all states of the second
nanocrystal are metastable, and only a shallow local minimum is formed
at a {\it larger} radius than the stable's one. Its critical size is 
also larger. These findings suggest that by necessity metastable
diamonds are on the average larger, {\it albeit} more deformed, than stable 
diamonds in a ta-C matrix would be. Again, we stress out that metastable
diamonds in a low-$z_{am}$ matrix would be sustained at moderately high
temperatures for practical mechanical purposes, but for a better
nanostructured material it is preferable to nucleate diamond cores
in a ta-C matrix.

\section{CONCLUSIONS}

In conclusion, we have presented in this paper a method to analyze
the stability of nanostructures embedded in an amorphous matrix.
This powerful and yet simple approach has yielded 
detailed information about the energetics and stability of na-C, and
gave insight into the interaction of nanocrystals with the
amorphous environment. Although the number of representative 
nanostructures could be extended to include other structures suggested
in the literature, we believe that the trends presented here are more 
general and encompass a wider range of situations relevant to na-C.

The central results of this analysis are: (a) only diamond and
schwarzite nanocrystals, among those who have been studied, are stable 
structures in an amorphous matrix, but not in the whole range of $z_{am}$.
The former are stable for $z_{am} \geq$ 3.6, the latter for 
$z_{am} \leq$ 2.9; (b) The other structures are metastable through the
whole range of $z_{am}$, but they have low formation energies at 
well-defined local minima, indicating that can be synthesized and 
sustained at moderate conditions for practical purposes; (c) There exists 
a possibility for transformation of the sp$^2$-bonded structures to 
metastable diamonds.

Future work includes the study of the electronic properties of na-C.
It is interesting to investigate whether nanocrystalline inclusions
in a-C alter its conductivity through the delocalization of the
$\pi$ states.\cite{Rober} Work toward this goal is in progress.
\section{ACKNOWLEDGMENTS}

We are grateful to Efthimios Kaxiras for granting us access to
the hardware resources of his group. 
This work was partly supported by a 
$\Pi$ENE$\Delta$ grant No. 99 E$\Delta$ 645, from the Greek General 
Secretariat for Research and Technology.

\newpage


\begin{center}
\begin{figure}
\includegraphics[width=0.7\textwidth]{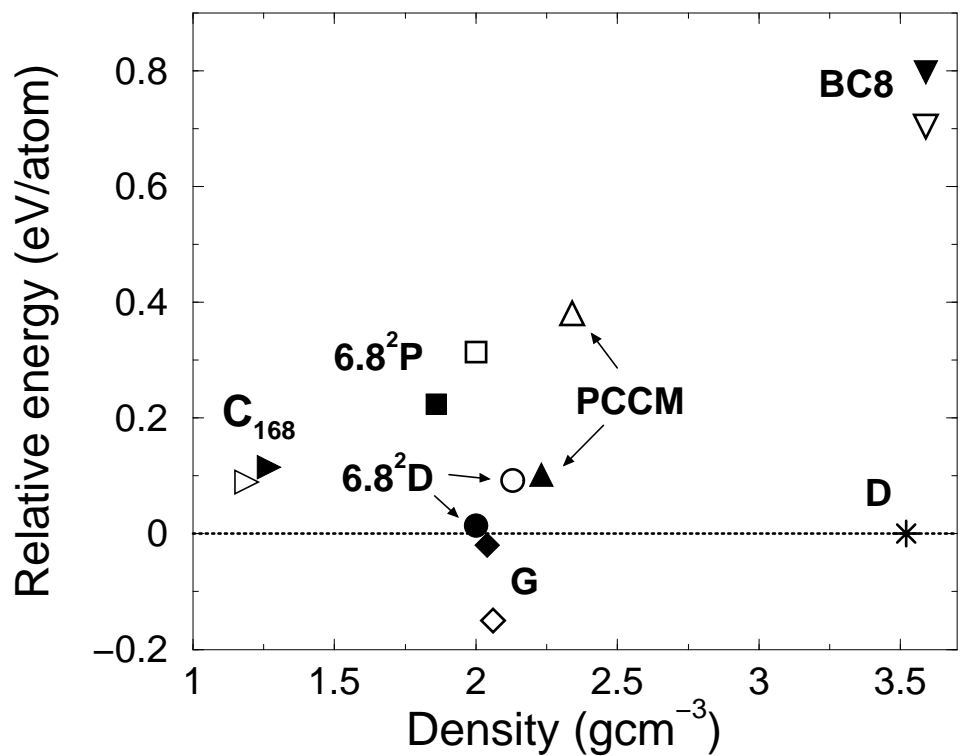}
\caption{Energies relative to diamond calculated for various polymorphs of 
carbon as a function of the density. Filled symbols show results with
the Tersoff potential, open symbols denote {\it ab initio} results.
G stands for graphite. (The empirical result is for a single layer).
Other abbreviations are explained in the text.}
\end{figure}

\vspace*{3cm}

\begin{figure}
\includegraphics[width=0.7\textwidth]{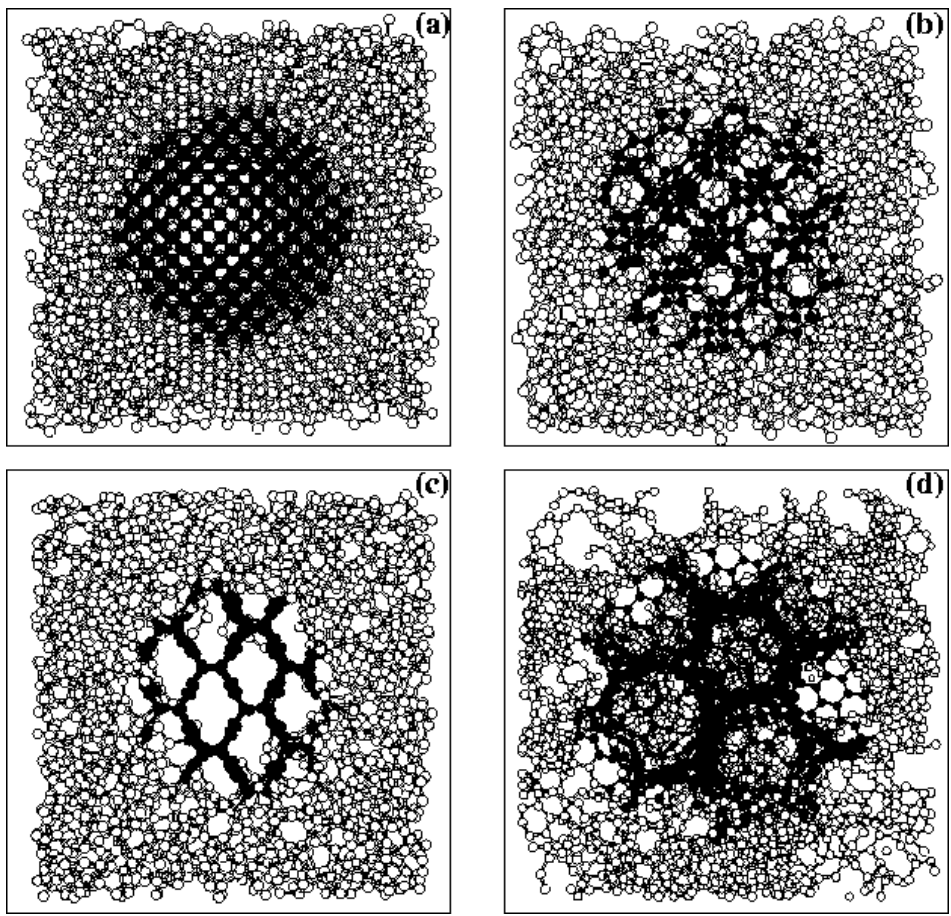}
\caption{Cuts from snapshots of four embedded nanostructures: (a) D; 
(b) 6.8$^2$D; (c) PCCM; (d) C$_{168}$. Open spheres denote atoms in the
amorphous matrix. Dark spheres show atoms in the nanocrystals. For 
clarity, several atoms in the amorphous matrix are not shown to reveal 
the structures.}
\end{figure} 

\vspace*{3cm}

\begin{figure}
\includegraphics[width=0.7\textwidth]{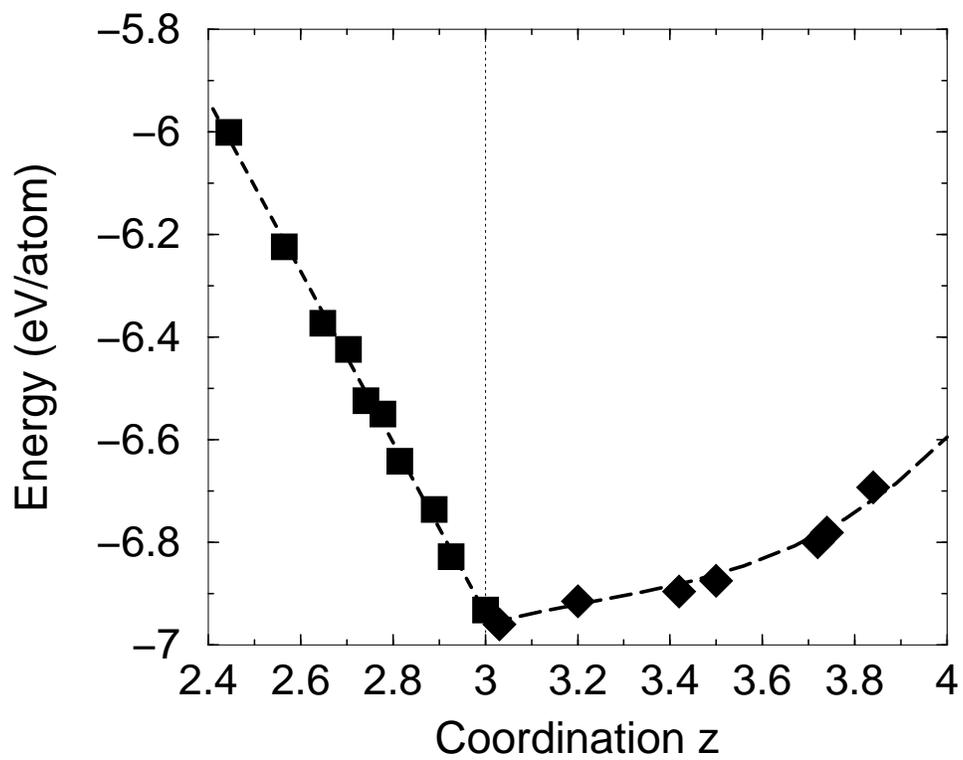}
\caption{Cohesive energy versus coordination for a-C configurations. Lines
are fits to the points.}
\end{figure} 

\vspace*{1cm}

\begin{figure}
\includegraphics[width=0.7\textwidth]{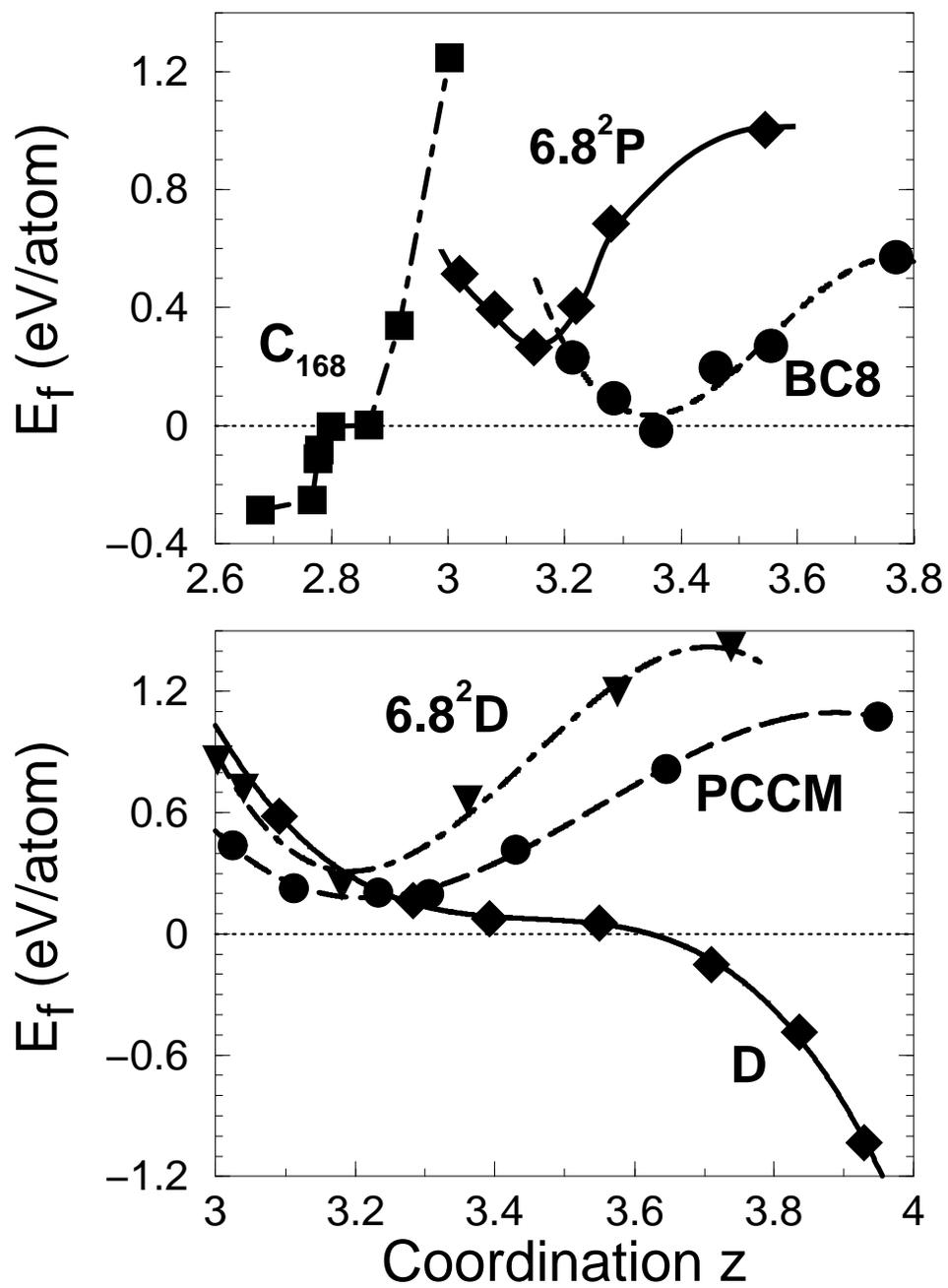}
\caption{Formation energies of the six nanostructures considered 
versus coordination of the amorphous matrix. Lines are fits to the points.}
\end{figure} 

\vspace*{3cm}

\begin{figure}
\includegraphics[width=0.7\textwidth]{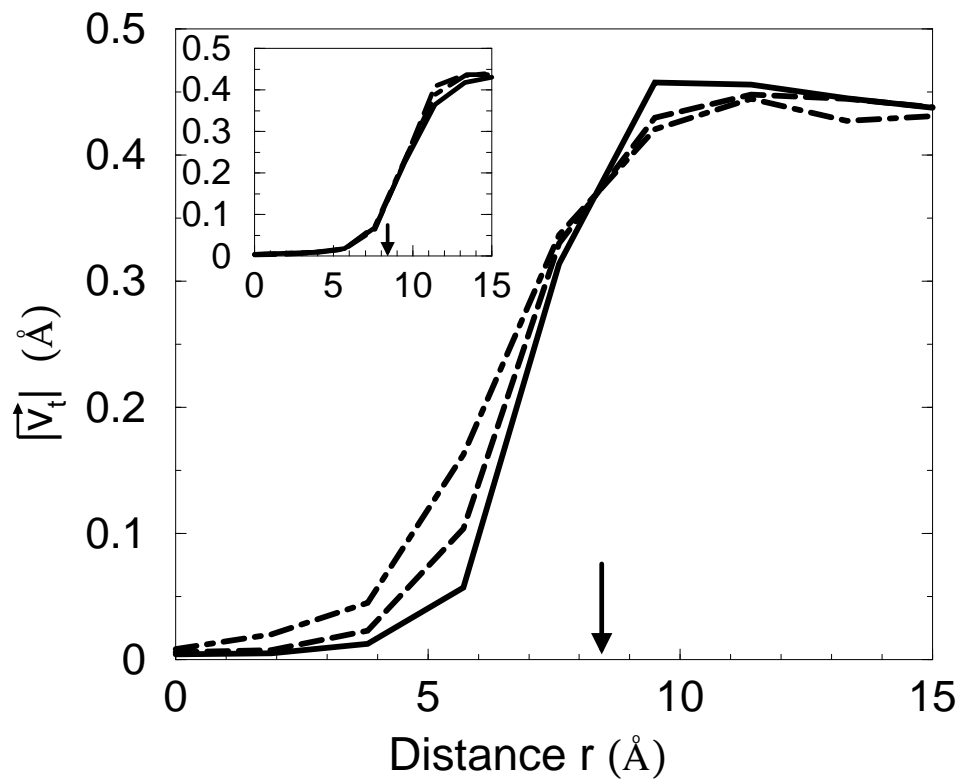}
\caption{Variation of the magnitude of the tetrahedral vector as a function
of distance from the center of diamond nanocrystals. A zero value
indicates ideal tetrahedral geometry. Main figure: metastable
case; inset: stable case. Arrows denote position of the interface.
Solid line: ``as grown''; dashed line: anneal at 1500 K; 
dash-dotted: anneal at 2000 K.}
\end{figure} 

\vspace*{3cm}

\begin{figure}
\includegraphics[width=0.7\textwidth]{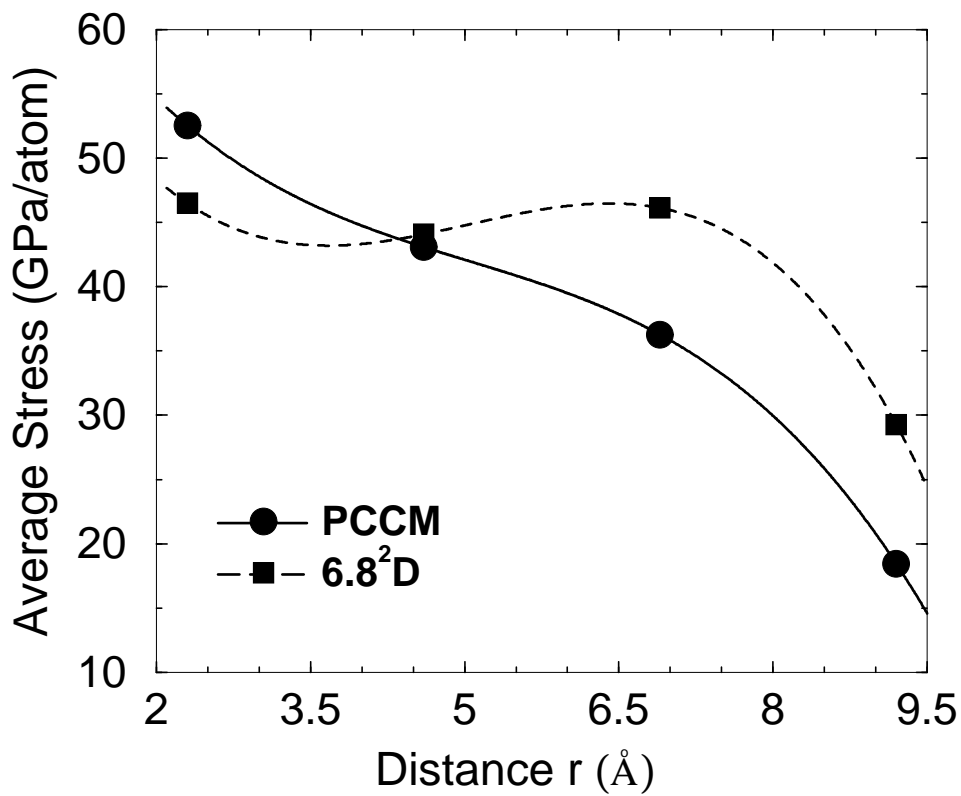}
\caption{Atomic stresses in PCCM and 6.8$^2$D nanocrystals, averaged over subsequent spherical shells of width 2.1 \AA, as a function of thedistance from the center. The stresses are calculated at 300 K.
In our formalism [13] positive values denote compressive
stresses. Lines are fits to the points.}
\end{figure} 
\vspace*{3cm}

\begin{figure}
\includegraphics[width=0.7\textwidth]{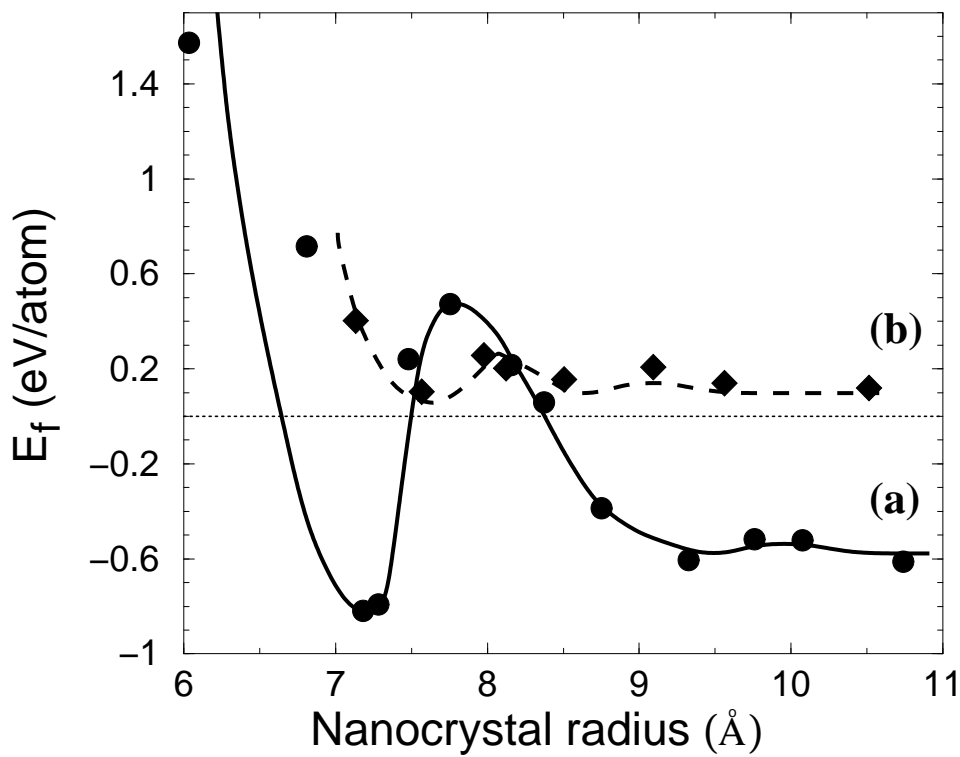}
\caption{Formation energy of (a) stable and (b) metastable diamond 
nanocrystals versus size.}
\end{figure} 
\end{center}

\end{document}